\documentclass[a4paper]{JHEP3} 
\usepackage{microtype}
\usepackage{lmodern}
\usepackage[T1]{fontenc}
\usepackage{color} 

\usepackage{graphicx}
\usepackage{dcolumn}
\usepackage{bm}
\usepackage{amssymb}
\usepackage{mathrsfs}
\usepackage{amsmath}
\usepackage{epsfig}

\usepackage{amsmath}
\usepackage{rotating}

\newcommand{\nova}{NO$\nu$A}
\newcommand{\dcp}{\delta_{CP}}
\newcommand{\cnv}{\v{C}erenkov}

\usepackage{epsf}
\def\be{\begin{equation}}
\def\ee{\end{equation}}
\def\bea{\begin{eqnarray}}
\def\eea{\end{eqnarray}}
\def\gsim{\ \rlap{\raise 2pt\hbox{$>$}}{\lower 2pt \hbox{$\sim$}}\ }
\def\lsim{\ \rlap{\raise 2pt\hbox{$<$}}{\lower 2pt \hbox{$\sim$}}\ }
\def\dslash{\kern-4pt \not{\hbox{\kern-2pt $\partial$}}}
\def\pslash{\not{\hbox{\kern-2pt p}}}

\title{Synergies between neutrino oscillation experiments: An `adequate' configuration 
for LBNO}
 \author{Monojit Ghosh, Pomita Ghoshal, Srubabati Goswami, Sushant K. Raut\\
 \llap{} Physical Research Laboratory, Navrangpura,
 Ahmedabad 380 009, India \\
 Email: \email{monojit@prl.res.in}, \email{pomita@prl.res.in}, 
 \email{sruba@prl.res.in}, \email{sushant@prl.res.in}
 }

\abstract
{
Determination of the neutrino mass hierarchy, octant of the mixing angle
$\theta_{23}$ and the CP violating phase $\dcp$ are the unsolved problems
in neutrino oscillation physics today. In this paper 
our aim is to obtain the minimum exposure required 
for the proposed Long Baseline
Neutrino Oscillation (LBNO) experiment to determine  
the above unknowns. 
We emphasize on the advantage of exploiting the  synergies
offered by the existing and upcoming long-baseline and atmospheric 
neutrino experiments
in economising the LBNO configuration.
In particular, we do a combined analysis for LBNO, T2K, \nova\ and INO. 
We consider three prospective 
LBNO setups -- CERN-Pyh\"{a}salmi ($2290$ km), CERN-Slanic
($1500$ km) and CERN-Fr\'{e}jus ($130$ km) and evaluate the 
adequate exposure required in each case.
Our analysis shows that   
the exposure required from LBNO can be reduced  
considerably
due to the synergies arising from the inclusion of the other experiments. 
}

\begin{document}
\maketitle


\section{Introduction}

The measurement of a non-zero $\theta_{13}$ by the reactor 
experiments Double Chooz \cite{dchooz_dec2011}, 
Daya bay \cite{dayabay_t13} and RENO \cite{reno_t13} 
is an important milestone in 
neutrino oscillation studies.
Together these experiments have given a more than $10\sigma$ significance 
in favour of a non-zero $\theta_{13}$ \cite{tortola2012,fogli2012,schwetz_nufit}. 
Recently a $7.5\sigma$ signal  for non-zero $\theta_{13}$ by
observing the $\nu_\mu - \nu_e$ oscillation has been announced  
by the 
T2K  experiment\cite{eps2013t2k}. 
The best-fit value of $\sin^2 2\theta_{13}$ as obtained from the 
global fits is close to 0.1. 
The discovery of a non-zero $\theta_{13}$ 
sets the stage for the determination of the   
remaining unknown neutrino oscillation parameters, namely --  
the ordering of neutrino mass eigenstates or mass hierarchy, the octant of 
the atmospheric mixing angle $\theta_{23}$ 
and the leptonic CP phase $\delta_{CP}$. This defines 
the road map for future programmes in neutrino oscillation physics.

The first set of  information on these quantities is expected to come from
the long-baseline (LBL) experiments T2K \cite{t2k} and \nova\ \cite{nova}.  
While T2K has already started operation and is giving results, \nova\ 
is scheduled to start data taking from 2014. 
Several studies have been carried out, exploring the 
potential of these experiments for determination of mass hierarchy,
octant and $\dcp$ 
\cite{degeneracy1,degeneracy3,twobase6,synergynt,cpcombo_sugiyama,menaparke,twobase1,hubercpv}. 
More recent studies in view  
of the measured value of $\theta_{13}$ can be seen in 
Refs.~\cite{novat2k,sanjib_glade,suprabhoctant,Blennow:2013swa,
octant_atmos,cpv_ino,minakata_cp}. 
The results obtained using the currently projected sensitivities 
for T2K and \nova\ can be summarized as follows:\\ 
(i)  Hierarchy can be determined at 95\% C.L. from  
the combined results from T2K and \nova\ for favourable values of 
$\delta_{CP}$. \\
(ii) Octant can be determined at 95\% C.L. by the T2K + \nova\ combination 
as long as $|45^\circ - \theta_{23}| > 6^\circ$ 
irrespective of hierarchy and $\delta_{CP}$.  
\\
(iii) Hint for a non-zero  $\delta_{CP}$ close to maximal CP violation 
can be obtained at 95\% C.L. This however requires a prior 
knowledge of mass hierarchy and octant of $\theta_{23}$. 

It was realized in \cite{cpv_ino} that although atmospheric neutrino 
experiments 
are insensitive to $\delta_{CP}$ themselves, they can play an important role 
in the detection of CP violation through their ability to determine 
mass hierarchy. The reason for hierarchy sensitivity of atmospheric 
neutrinos can be attributed to the large matter effects experienced
by the neutrinos while passing through longer path lengths en route the 
detector \cite{atmoshier1,atmoshier2,atmoshier3}. 
The major future atmospheric neutrino projects are HyperKamiokande 
and MEMPHYS using water \cnv\ technology \cite{hkloi,campagne}, 
India-based Neutrino Observatory (INO)
which will be using a magnetized 
iron calorimeter detector \cite{inowebsite} 
and PINGU which is an upgraded version of the 
IceCube detector and will use Antarctic ice as detector material 
and strings of digital optical modules as the detector element \cite{pingu}.
Large volume liquid Argon detectors have also been proposed \cite{lar1,raj}.
The capabilities of these experiments have been investigated 
in detail in several recent papers, see for example 
Refs.~\cite{schwetzblennow,gct,tarak,Ghosh:2013mga,pingusmirnov,pinguagarwalla,pinguwinter}. 
In particular, the synergy between the LBL experiments and INO 
for determination of mass hierarchy has been discussed in 
\cite{schwetzblennow,gct},  
for octant determination has been explored in 
\cite{octant_atmos} and that for $\dcp$ has been studied in
\cite{cpv_ino}. 
The reason for this synergy lies in the different 
baselines, neutrino energy, earth matter effects and source and detector 
characteristics involved in various long-baseline and 
atmospheric experiments. 
This leads to a different dependence of their oscillation probabilities 
on the parameters  making their data complementary to each other, 
increasing the sensitivity. 
However from the results obtained in the above studies 
one concludes that even if the current LBL experiments T2K and \nova\ 
join forces with INO (which has already been granted project approval
\cite{nkm_nufact}), a 
conclusive 5$\sigma$ evidence  for the unknown parameters would 
still require new experiments. 

One of the promising proposals for an oscillation experiment beyond 
the current and upcoming ones, is the LAGUNA-LBNO project in Europe
\footnote{The LBNE collaboration in US is also considering the 
same physics goals \cite{lbne,raj}.}.   
The source of neutrinos for this experiment is likely to be at CERN. 
Various potential sites for the detector 
have been identified by LAGUNA, including Boulby (U.K.), 
Canfranc (Spain), Fr\'{e}jus (France), Pyh\"{a}salmi (Finland), Slanic 
(Romania), SUNLAB (Poland) and Umbria (Italy) \cite{laguna_options}. 
Previous studies 
have already shown that some of these potential experiments can have very 
good capability for measuring the unknown parameters
\cite{laguna_options,lbno_eoi,incremental}. 
However, the precise 
configuration of LBNO is currently being deliberated 
and it is desirable to adjudge the information that can be gleaned 
from the combination of  current generation 
LBL+atmospheric experiments in the planning of 
this experiment. In this paper we embark on such an exercise. 
We consider the iron calorimeter (ICAL) detector proposed 
by the INO collaboration as the atmospheric detector in conjunction
with the LBL experiments T2K and \nova\ and 
determine the configuration for LBNO with 
`adequate' exposure
which can determine the unknown 
oscillation parameters.
The `adequate' configuration is defined as one with the minimal exposure
which would give a 5$\sigma$ discovery potential for hierarchy 
and octant and 3$\sigma$ discovery potential for $\dcp$ 
in the most unfavourable case. 
This configuration can be viewed as 
the first step in a staged approach that has been advocated by previous 
studies \cite{incremental}. 
 
The plan of this paper is as follows. In the next section we give the 
experimental specifications that we have used to simulate \nova,
T2K, INO and the proposed LBNO experiment. We then discuss briefly 
the synergies between neutrino oscillation experiments. 
The next three 
sections thereafter are devoted to 
the analysis of the experimental reach of the combination of experiments 
for determining the mass hierarchy, octant of $\theta_{23}$ and CP violation 
respectively. Finally, we summarize our results.

\section{Simulation details}

In this paper, we have considered the contributions of \nova, T2K, ICAL@INO and 
LBNO towards determining the mass hierarchy, octant of $\theta_{23}$ and 
CP violation. 
Simulations of all long-baseline experiments were carried out 
using the GLoBES package\cite{globes1,globes2} along with 
its auxiliary data files \cite{messier_xsec,paschos_xsec}. 
Given below are the specifications 
of these experiments. 

For \nova\ and T2K, we have considered the standard detector and beam 
specifications used in Ref.~\cite{sanjib_glade}. 
\nova, with $7.3\times10^{20}$ protons on target (pot) per year is assumed 
to run for $3$ years each in neutrino and antineutrino mode. The neutrinos 
are detected at a $14$ kt TASD detector placed $14$ mrad off-axis, at a distance 
of $812$ km from the NuMI source. We have used the new efficiencies and 
resolutions for \nova\ which are optimized for the moderately large 
value of $\theta_{13}$ \cite{sanjib_glade,Kyoto2012nova}. For T2K, 
the current plan is 
to have a total of $7.8\times10^{21}$ pot over the entire runtime of T2K. 
In our 
simulations, we have adjusted the runtime so as to get a total of 
$\sim 8\times10^{21}$ pot. 
In this work, we have 
assumed that T2K will run entirely with neutrinos. 
We have taken a baseline of $295$ km 
and detector mass of $22.5$ kt for this experiment. 
The relevant experimental 
specifications have been taken from Refs.~\cite{hubercpv,t2k,globes_t2k4,globes_t2k5}.

The ICAL detector at the INO site in southern India is a $50$ kt magnetized 
iron calorimeter, which will detect muon neutrino events with the capacity for 
charge detection provided by a magnetic field of about $1.3$ tesla. Charge 
identification allows a separation of neutrino and antineutrino events,
which is advantageous for mass hierarchy determination. The detector is under construction 
and is expected to start functioning within a projected time frame of about 5 years. 
We have considered a $10$ year run for this atmospheric 
neutrino experiment, giving it a total exposure of $500$ kt yr. 
The neutrino energy and angular resolution of the detector are taken to be 
$0.1\sqrt{E(\textrm{GeV})}$ and $10^\circ$ respectively, while its 
efficiency is taken to be $85\%$. 
These effective resolutions and efficiencies give 
results comparable to those obtained through a full detector simulation
\cite{gct}. 

Out of the various possible options for the LBNO experiment listed in the 
previous section, we consider the following three options that are 
prominent in the literature: 
CERN-Pyh\"{a}salmi, CERN-Slanic and CERN-Fr\'{e}jus 
\footnote{The three options shortlisted for LBNO are CERN-Pyh\"{a}salmi, 
CERN-Umbria and CERN-Fr\'{e}jus. However, we consider the CERN-Slanic option 
(1540 km baseline), because it lies in the range of baselines that are well 
suited for $\dcp$ studies \cite{kopphuber_2base}. 
Thus, the three setups analyzed in this paper 
cover the full range of baselines under consideration.}. 
The specifications that we have used in this work are listed below in 
Table~\ref{tab:exp_lbno}. We have used the superbeam fluxes from 
Ref.~\cite{lbnoflux}. 
We have explicitly taken into account the effect of wrong-sign 
contamination for these experiments. In particular, we find that the neutrino 
contamination in the antineutrino beam can have a significant effect on the 
event rates. 

\begin{table}[htb]
\begin{center}
 
     \begin{tabular}{ || l || c | c | c ||}

         \hline
         \hline
         Detector site & Pyh\"{a}salmi & Slanic & Fr\'{e}jus \\
         \hline
         \hline
         Baseline & $2290$ km & $1540$ km & $130$ km \\
         Detector Type & LArTPC & LArTPC & Water \v{C}erenkov \\
	 Proton energy & $50$ GeV & $50$ GeV & $4.5$ GeV \\
         Resolutions, efficiencies & as in Ref.~\cite{incremental} &  
			as in Ref.~\cite{incremental} &  as in Ref.~\cite{newmemphys} \\
	 Signal systematics & 5\% & 5\% & 5\%\\
         Background systematics & 5\% & 5\%  & 10\% \\
         \hline
         \hline
      \end{tabular}
         
\caption{\footnotesize Experimental characteristics of the LBNO options 
considered in this paper.}

\label{tab:exp_lbno}

\end{center}
\end{table}

We have fixed the `true' values of the parameters close to the values obtained
from global fits of world neutrino data \cite{tortola2012,fogli2012,schwetz_nufit}. 
We have taken: 
$\sin^2 \theta_{12} = 0.304$, $|\Delta_{31}| = 2.4 \times 10^{-3}$ eV$^2$,
$\Delta_{21} = 7.65 \times 10^{-5}$ eV$^2$ and $\sin^2 2\theta_{13} = 0.1$. 
Three representative
true values of $\theta_{23}$ have been considered -- $39^\circ$, 
$45^\circ$ and $51^\circ$ (except in the case of octant determination where 
a wider range and more intermediate values have been included). The true 
value of $\dcp$ is varied in its entire allowed range. All our 
results are shown for both cases -- normal hierarchy (NH):
$m_1 < m_2 <<m_3$
and inverted 
hierarchy (IH):  $m_3 \ll m_1 \lsim m_2$. The
`test' values of the parameters are allowed to vary in the following ranges -- 
$\theta_{23} \in \left[35^\circ,55^\circ\right]$, 
$\sin^2 2\theta_{13} \in \left[0.085,0.115\right]$, 
$\dcp \in \left[0,360^\circ\right)$. The test hierarchy is also allowed 
to run over both possibilities. 
We have imposed a prior 
on the value of $\sin^2 2\theta_{13}$ with an error 
$\sigma(\sin^2 2\theta_{13}) = 0.005$, which is the expected precision on this 
parameter from the reactor neutrino experiments \cite{daya_005}. 
We have however not imposed any prior on the atmospheric parameters, 
instead allowing the $\nu_\mu$ disappearance channels to restrict 
their range. In all our simulations, we 
have taken into account the three-flavour-corrected definitions of the 
atmospheric parameters \cite{parke_defn,degouvea_defn,spuriousth23}. 

In the following sections, we analyze the ability of the experiments 
\nova, T2K, ICAL@INO and LBNO to collectively determine the neutrino mass 
hierarchy, octant of $\theta_{23}$ and detect CP violation. We demand that 
this combination of experiments determine the mass hierarchy and octant of 
$\theta_{23}$ with a statistical significance corresponding to $\chi^2=25$, 
and that CP violation be detected with $\chi^2=9$ 
\footnote{Conventionally, these 
values are taken to correspond to $5\sigma$ and $3\sigma$, respectively. 
However, it was recently pointed out in Refs.~\cite{stats_hier1,stats_hier2} 
that for a binary question such as hierarchy, the relation between $\chi^2$ 
and confidence levels is somewhat involved. For more recent discussions on 
statistical interpretation, see Refs.~\cite{blennowstats1,blennowstats2,lbno2013dec}}. 

The aim of this exercise is to determine the least exposure required from LBNO 
in order to fulfil the above demands. 
Therefore, we have plotted the sensitivity 
to hierarchy/octant/CP violation for various different exposures of LBNO, 
combined with \nova, T2K and INO. 
From this, we estimate the adequate amount 
of exposure required by LBNO. We express the exposure in units of pot-kt. 
This is a product of three experimental quantities: 
\begin{equation}
 \textrm{exposure (pot-kt)} = \textrm{beam intensity (pot/yr)} \times 
 \textrm{runtime (yr)} \times \textrm{detector mass (kt)} ~.
\end{equation}
Thus, a given value of exposure can be achieved experimentally by 
adjusting the intensity, runtime and detector mass. 
The advantage of using this 
measure is that while the physics goals are expressed in terms of simply 
one number (the exposure), the experimental implementation of this exposure 
can be attained by various combinations of beam, detector and runtime settings. 
For example, an exposure of $45\times10^{21}$ pot-kt could be achieved with 
a $1.5\times10^{21}$ pot/yr beam running for $3$ years 
with a $10$ kt detector or a $3 \times 10^{21}$ pot/yr beam running for 
$3$ years with a $5$ kt detector. 
In the terminology used in this paper,
the exposures given correspond to each 
mode (neutrino and antineutrino). 
Thus, 
a runtime of $n$ years implies $n$ years 
each in neutrino and antineutrino mode totalling to $2n$ years.

\section{Synergies between oscillation experiments}

\begin{figure}[htb]
\begin{tabular}{rl}
{\footnotesize 1540 km, NH (hierarchy)} \hspace{2cm} & \hspace{3cm}  {\footnotesize 1540 km, IH (octant)} \\
\epsfig{file=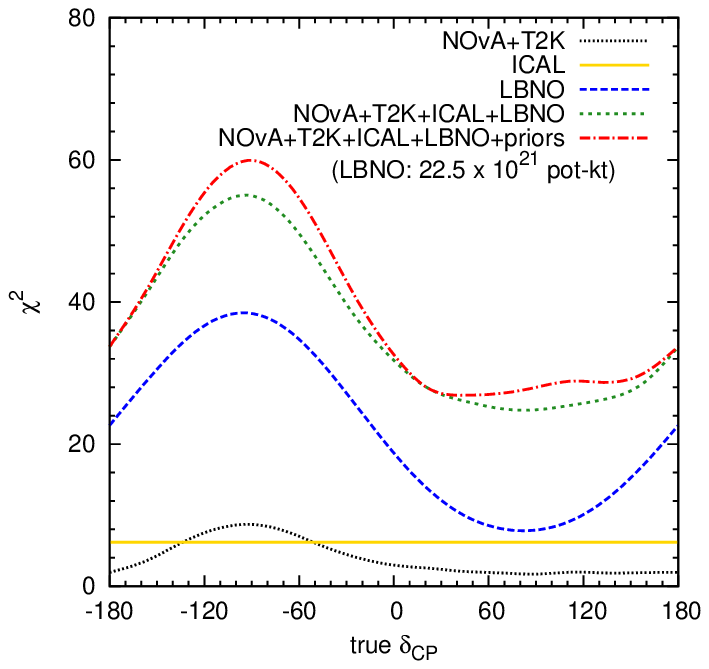, bbllx=85, bblly=50, bburx=300, bbury=250,clip=} &
\epsfig{file=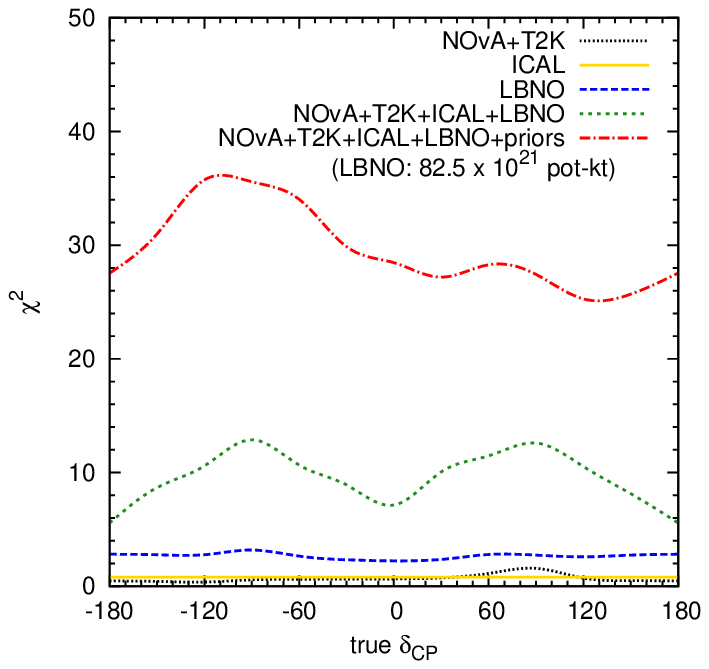, bbllx=85, bblly=50, bburx=300, bbury=250,clip=}
\end{tabular}
\caption{\footnotesize Demonstration of synergy between various oscillation experiments, 
by plotting hierarchy(octant) sensitivity $\chi^2$ vs true $\dcp$ in the left(right) 
panel. The curves are without priors unless specified. It is clear that the combined 
$\chi^2$ is greater than the sum of individual $\chi^2$ values.
}
\label{fig:synergy}
\end{figure}

Neutrino oscillation parameters are measured by observing events at a detector, and 
inferring the oscillation probability from them. In different experiments (and oscillation 
channels), neutrinos travel different distances and have different energies. Moreover, 
depending on the baseline, they experience matter effects to varying degrees. The 
energy spectrum of the events seen at the detector is also affected by the initial 
flux of neutrinos. As a result of these effects, the dependence of the event spectrum 
on the oscillation parameters can be different. 

When we try to fit the events to a set of oscillation parameters, data from various 
experiments tend to choose slightly different best-fit points. This was demonstrated 
explicitly in the context of octant sensitivity in Ref.~\cite{octant_atmos}. In a 
combined fit, data from each experiment gives (in general) some $\chi^2$ at the 
best-fit point of the other experiments. As a result, the net $\chi^2$ of a combined 
analysis is greater than the sum of the individual minima. Therefore, we say that 
there is a synergy between various experiments. This is the very 
principle that leads to the lifting of parameter degeneracies using various experiments. 

In the left(right) panel of Fig.~\ref{fig:synergy}, we have shown the 
hierarchy(octant) determination capability 
of various experiments separately (without including priors) as well as from 
their combined analysis (without and with priors) for true $\theta_{23}=39^\circ$. 
In the left(right) panel, for LBNO, we have used the $1540$ km setup with an exposure of 
$22.5(82.5)\times10^{21}$ pot-kt, and assuming NH(IH) to be the true hierarchy.
It is clear to see that the combined $\chi^2$ is much larger than the sum of the 
individual contributions. 
For hierarchy determination, the effect of synergy is more pronounced around 
$\dcp=90^\circ$ where the effect of degeneracy is maximum. For more favourable values 
of $\dcp$, the effect is milder. 
In the plot for octant sensitivity, we find that apart from the synergy between 
long-baseline and atmospheric neutrino experiments, there is a tremendous synergy 
between these and the reactor neutrino data. This is evident from the substantial effect 
of adding the $\theta_{13}$ prior. 
The synergy between experiments for octant sensitivity is discussed 
in detail in Ref.~\cite{octant_atmos}. The synergy between long-baseline and 
atmospheric neutrino experiments in detecting CP violation has been pointed out in 
Ref.~\cite{cpv_ino,ourlongcp}.

\section{Determination of mass hierarchy}

\begin{figure}[htb]
\begin{tabular}{rl}
{\footnotesize 2290 km, NH} \hspace{2cm} & \hspace{3cm}  {\footnotesize 2290 km, IH} \\
\epsfig{file=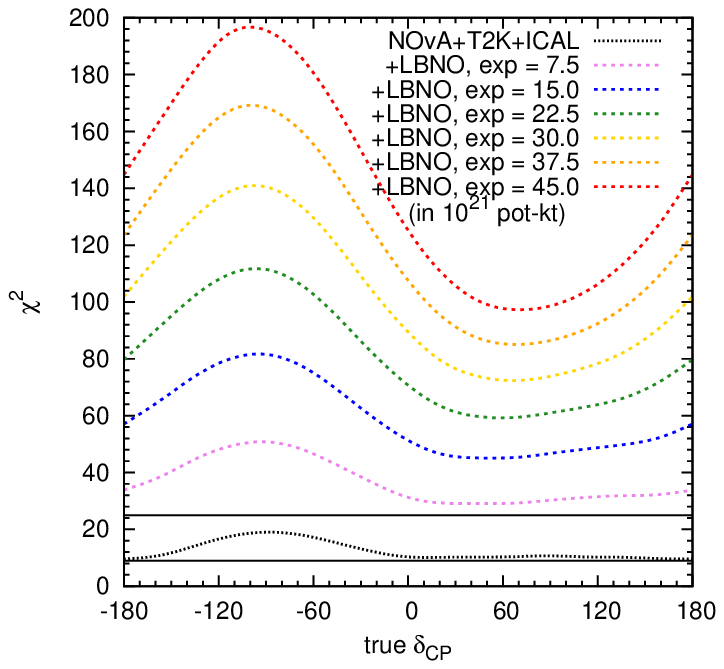,bbllx=85, bblly=50, bburx=300, bbury=250,clip=} &
\epsfig{file=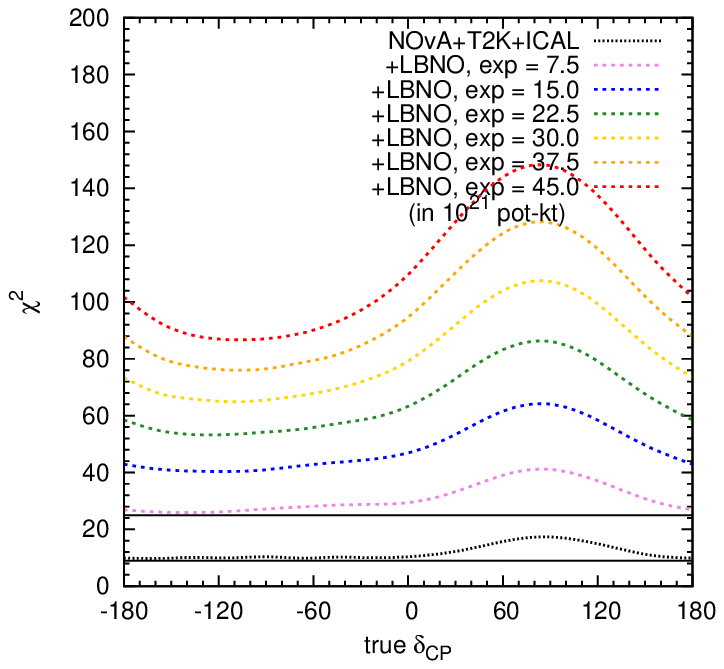,bbllx=85, bblly=50, bburx=300, bbury=250,clip=} \\
{\footnotesize 1540 km, NH} \hspace{2cm} & \hspace{3cm} {\footnotesize 1540 km, IH} \\
\epsfig{file=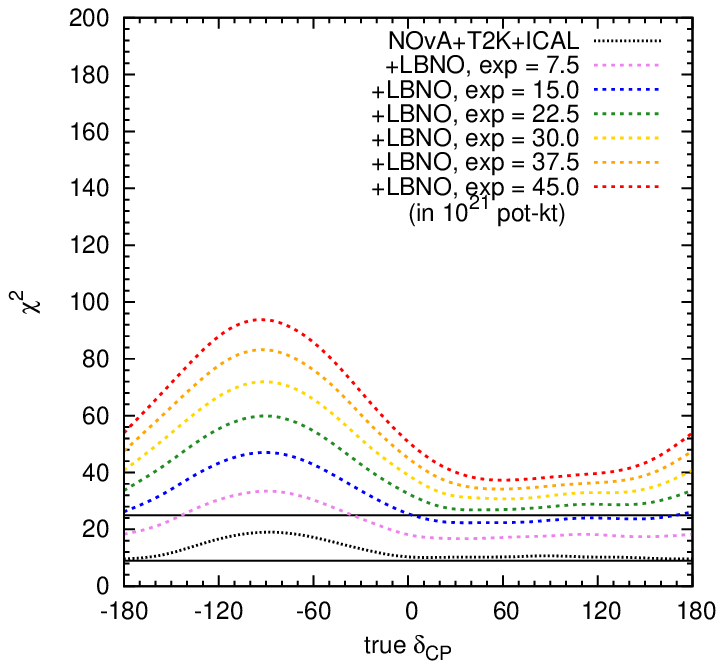,bbllx=85, bblly=50, bburx=300, bbury=250,clip=} &
\epsfig{file=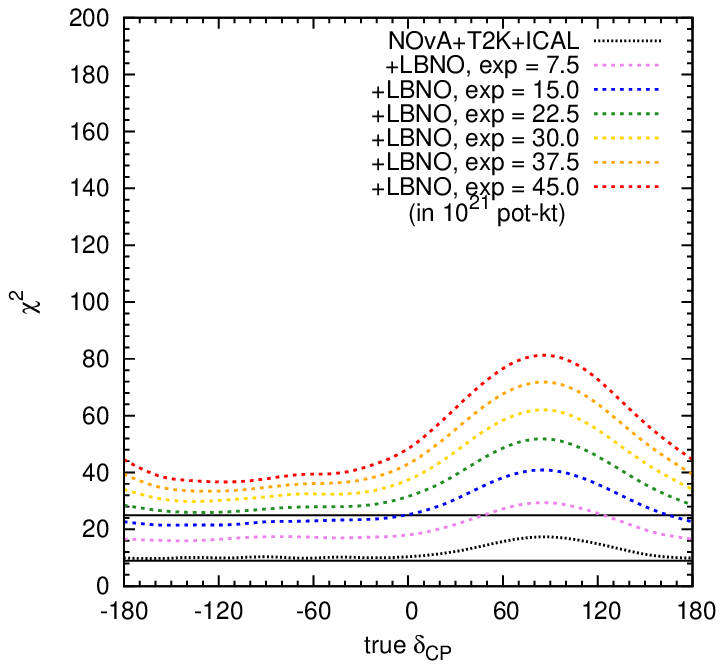,bbllx=85, bblly=50, bburx=300, bbury=250,clip=}
\end{tabular}
\caption{\footnotesize Hierarchy sensitivity $\chi^2$ vs true $\dcp$. 
The top(bottom) panels are for the $2290$($1540$) km baseline. The left(right) 
panels are for true NH(IH). In all the panels, the lowermost densely-dotted (black) 
curve is for \nova+T2K+ICAL, while the curves above are for \nova+T2K+ICAL+LBNO, 
for various values of LBNO exposure. All the plotted sensitivities are for the least 
favourable value of true $\theta_{23}$.}
\label{fig:hierall}
\end{figure}

\begin{figure}[htb]
\begin{center}
\epsfig{file=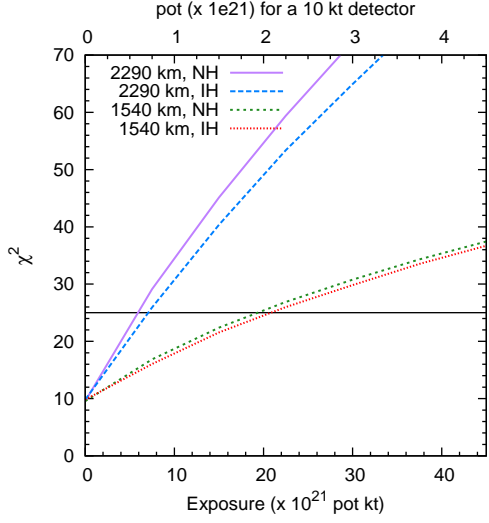, bbllx=85, bblly=50, bburx=300, bbury=255,clip=}
\end{center}
\caption{\footnotesize Hierarchy sensitivity $\chi^2$ vs LBNO 
exposure, for both baselines and hierarchies under consideration. 
The value of exposure shown here is adequate to exclude the wrong hierarchy 
for all values of $\dcp$. 
The additional 
axis along the upper edge of the graph shows the required total pot assuming a 
detector mass of $10$ kt.}
\label{fig:hierexpo}
\end{figure}

\begin{figure}[htb]
\begin{tabular}{rl}
\epsfig{file=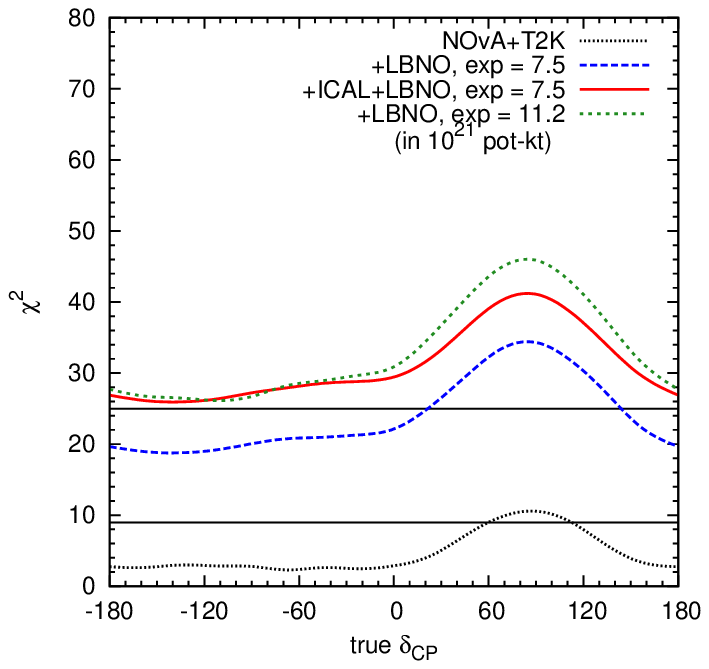, bbllx=85, bblly=50, bburx=300, bbury=250,clip=} &
\epsfig{file=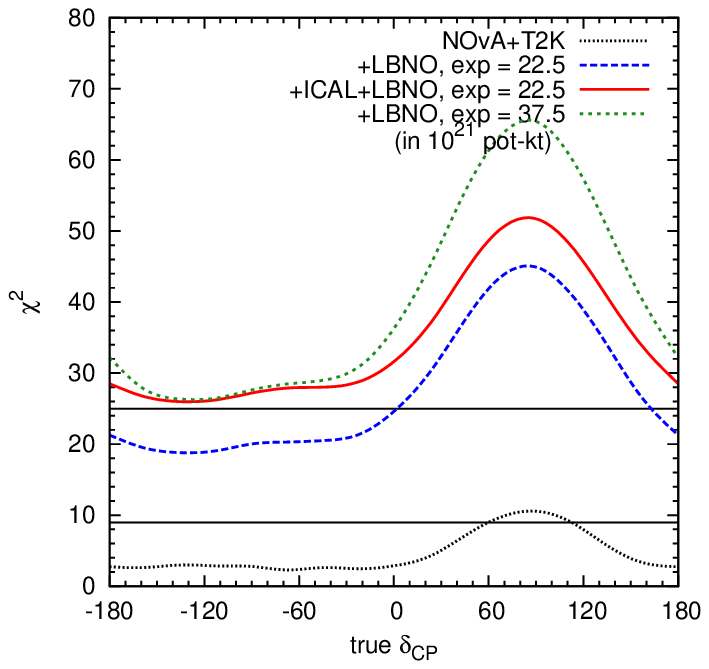, bbllx=85, bblly=50, bburx=300, bbury=250,clip=}
\end{tabular}
\caption{\footnotesize Hierarchy sensitivity $\chi^2$ for different 
 combinations of experiments, demonstrating the synergy between them. The left(right) 
 panel is 
 for a LBNO baseline of $2290$($1540$) km, assuming IH to be true. With only 
 T2K+\nova+LBNO 
 (dashed, blue), the sensitivity is lower than for T2K+\nova+LBNO+ICAL (red, solid).
 Without ICAL data, the LBNO exposure would have to be increased substantially 
 (dotted, green) in order to get comparable sensitivity. All the plotted sensitivities 
 are for the least favourable value of true $\theta_{23}$.}
\label{fig:hiersyn}
\end{figure}

Long-baseline experiments such as \nova\ and T2K primarily use the 
$\nu_\mu \rightarrow \nu_e$ 
oscillation channel $P_{\mu e}$ to determine the neutrino mass hierarchy. 
Using the approximate perturbative formula for this probability
\cite{cervera,freund,akhmedov}, 
it can be seen that there is a hierarchy-$\dcp$ degeneracy \cite{degeneracy2}. 
As a result, the 
hierarchy sensitivity of these experiments is a strong function of the value of 
$\dcp$ in nature. In Refs.~\cite{novat2k,degeneracy2}, it was shown that 
there exist favourable and 
unfavourable combinations of hierarchy and $\dcp$ for the hierarchy 
sensitivity of LBL experiments.
Combining information from \nova\ and T2K improves 
the hierarchy sensitivity in the unfavourable part of the parameter space. 

On the other hand, the hierarchy sensitivity of an atmospheric neutrino 
experiment 
like ICAL is almost independent of $\dcp$. This is due to the effect of angular 
smearing that washes out the $\dcp$-dependence \cite{cpv_ino}. Therefore, 
irrespective of the value of $\dcp$ in nature, ICAL can determine the 
mass hierarchy. Thus combining ICAL results with that of 
T2K and \nova\ is expected to give an enhanced 
sensitivity to mass hierarchy 
independently of the value of $\dcp$ \cite{schwetzblennow,gct}. 

Among the three chosen prospective baselines for LBNO, the 130 km 
setup has the lowest hierarchy sensitivity due to small matter effects. 
As the baseline increases, the hierarchy sensitivity becomes better because 
of enhanced matter effects. In particular, the 2290 km setup has the 
unique advantage of being close to satisfying the bimagic conditions 
\cite{bnlhs,bnlhs_long,bimagic}. 
This feature makes the baseline particularly suited for 
hierarchy determination.  The above features are reflected in Fig.~\ref{fig:hierall}.
In each of the panels of Fig.~\ref{fig:hierall}, 
the lowermost densely-dotted (black) 
curve shows the hierarchy sensitivity of the combination \nova+T2K+ICAL. 
We see that these experiments can collectively give $\chi^2 \approx 9$ 
sensitivity to 
the hierarchy. Therefore, in keeping with our aims, we need to determine 
the minimum exposure for LBNO, such that the combination \nova+T2K+ICAL+LBNO 
crosses the threshold of $\chi^2=25$ for all values of $\dcp$. For this, 
we have plotted the combined sensitivity of \nova+T2K+ICAL+LBNO for 
various values of LBNO exposure 
(in units of 
$10^{21}$ pot-kt). The results are shown for two baselines -- $2290$ km and 
$1540$ km, and for both hierarchies. 
We find that our results are consistent with 
those shown in Ref.~\cite{lbno_eoi}, for the same beam power and oscillation 
parameters.
For the baseline of $130$ km, 
it is not possible to cross $\chi^2=25$ even with extremely high 
exposure. Therefore we have not shown 
the corresponding plots for this baseline. 
We considered three true values of 
$\theta_{23}$ -- $39^\circ, 45^\circ, 51^\circ$ and chose the 
least favourable  of these in generating the figures.  
Thus, our results represent the most conservative case.  
We find that in most cases, the minimum $\chi^2$ for hierarchy 
determination occurs for true $\theta_{23} = 39^\circ$. 

Finally, in Fig.~\ref{fig:hierexpo}, we have 
condensed all this 
information into a single plot. We have shown the sensitivity for the 
experiments as a function of the LBNO exposure. We see that for $2290$($1540$) 
km, it is sufficient for LBNO to have an exposure of around 
$7\times10^{21}$($21\times10^{21}$) pot-kt in order to get $\chi^2=25$ sensitivity 
for all values of $\dcp$. Along the upper edge of the graph, 
we have provided an additional axis, which denotes the total pot required 
if we assume that the detector has a mass of $10$ kt. For $2290$($1540$) km, 
we need a total of $0.7\times10^{21}$($2.1\times10^{21}$) pot.
To get some idea of the time scale involved we consider for instance the 
beam intensity used in Ref.~\cite{incremental} which corresponds to 
$3 \times 10^{21}$ pot/yr delivered by a $50$ GeV proton beam 
from CERN with beam power $1.6$ MW. 
The total pot of $0.7\times10^{21}$ for a 10 kt detector at the $2290$ 
($1540$) km 
baseline  would thus need 
less than $1$($2$) years (total, inclusive of $\nu$ and $\overline{\nu}$ 
runs) to establish mass hierarchy with $\chi^2=25$. 

Fig.~\ref{fig:hiersyn}, demonstrates the synergy 
between long-baseline and atmospheric neutrino experiments. We have chosen 
the $2290$($1540$) km baseline as an illustrative case
in the left(right) panels,
with the true hierarchy 
assumed to be IH. The densely-dotted (black) curve at the bottom shows 
the hierarchy sensitivity of \nova+T2K 
without any atmospheric neutrino data included in the analysis.
If the atmospheric information is not included then the combination 
of \nova+T2K+LBNO would need about $11  \times10^{21}$ pot-kt in order to
attain $\chi^2=25$, for the 2290 baseline. 
Assuming a beam intensity of $3 \times 10^{21}$ pot/yr 
this would require less than a year to measure the 
hierarchy with a 10 kt detector. 
Combining these with ICAL reduces the exposure to 
$7 \times10^{21}$ pot-kt. Thus, 
for the same beam intensity one can achieve the same sensitivity
with a 7 kt detector.  
Similar conclusions can be drawn for the 1540 km set-up. 
It should be noted that the numbers in Fig.~\ref{fig:hiersyn} are 
sample values at which the simulations are performed. 
The exposure required for each set-up
to attain the `adequate' values can be read off from  
Fig.~\ref{fig:hierexpo} and is presented in Table
\ref{tab:res}. 


\section{Determination of octant of $\theta_{23}$}

\begin{figure}[htb]
\begin{tabular}{rl}
{\footnotesize 2290 km, NH} \hspace{2cm} & \hspace{3cm}  {\footnotesize 2290 km, IH} \\
\epsfig{file=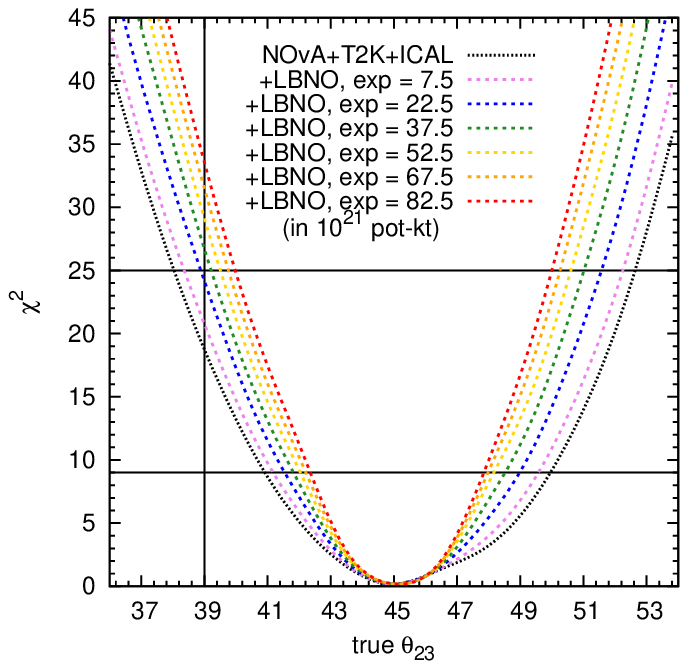,bbllx=85, bblly=50, bburx=300, bbury=250,clip=} &
\epsfig{file=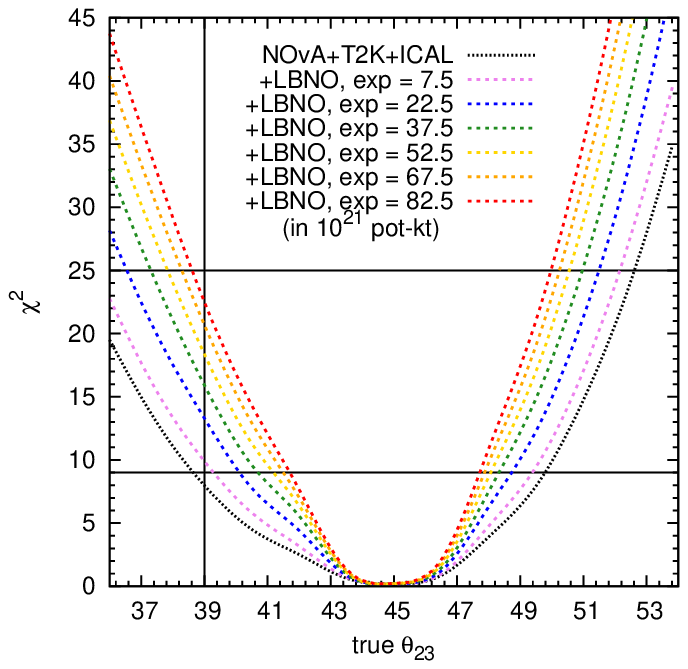,bbllx=85, bblly=50, bburx=300, bbury=250,clip=} \\
{\footnotesize 1540 km, NH} \hspace{2cm} & \hspace{3cm} {\footnotesize 1540 km, IH} \\
\epsfig{file=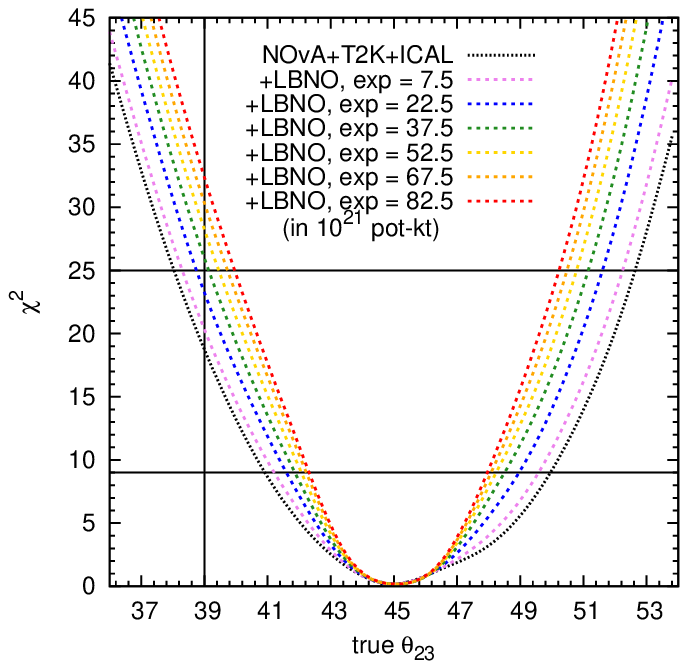,bbllx=85, bblly=50, bburx=300, bbury=250,clip=} &
\epsfig{file=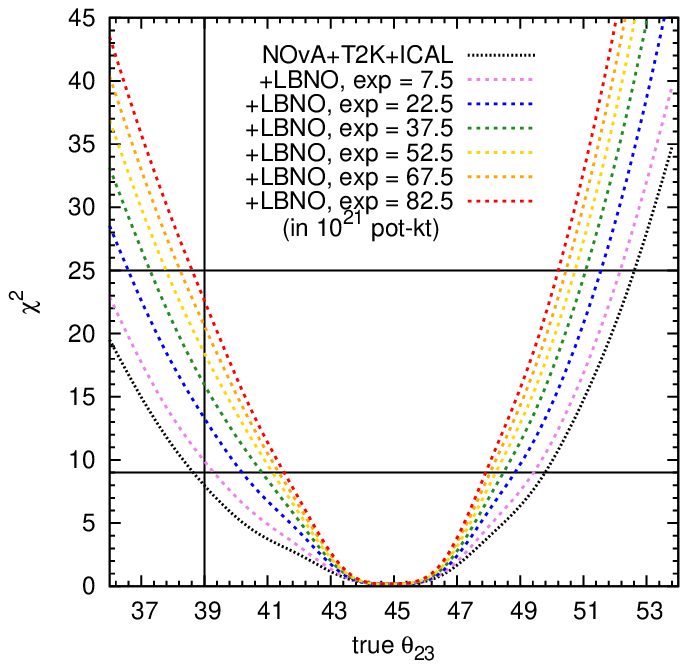,bbllx=85, bblly=50, bburx=300, bbury=250,clip=}
\end{tabular}
\caption{\footnotesize Octant sensitivity $\chi^2$ vs true $\theta_{23}$. 
The top(bottom) panels are for the $2290$($1540$) km baseline. The left(right) 
panels are for true NH(IH). In all the panels, the lowermost densely-dotted (black) 
curve is for \nova+T2K+ICAL, while the curves above are for \nova+T2K+ICAL+LBNO, 
for various values of LBNO exposure. All the plotted sensitivities 
 are for the least favourable value of true $\dcp$.}
\label{fig:octall}
\end{figure}

\begin{figure}[htb]
\begin{center}
\epsfig{file=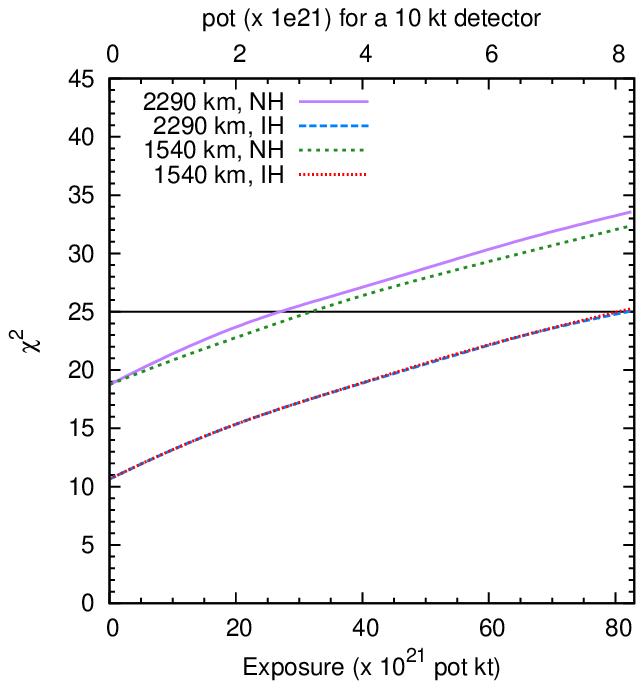, bbllx=85, bblly=50, bburx=300, bbury=255,clip=}
\end{center}
\caption{\footnotesize Octant sensitivity $\chi^2$ vs LBNO 
exposure, for the $2290$ km and $1540$ km baselines and both hierarchies, with 
$\theta_{23}=39^\circ$. The additional axis along 
the upper edge of the graph shows the required total pot assuming a detector 
mass of $10$ kt.}
\label{fig:octexpo}
\end{figure}

\begin{figure}[htb]
 \begin{tabular}{rl}
  {\footnotesize 130 km, NH} \hspace{2cm} & \hspace{3cm} {\footnotesize 130 km, IH} \\
\epsfig{file=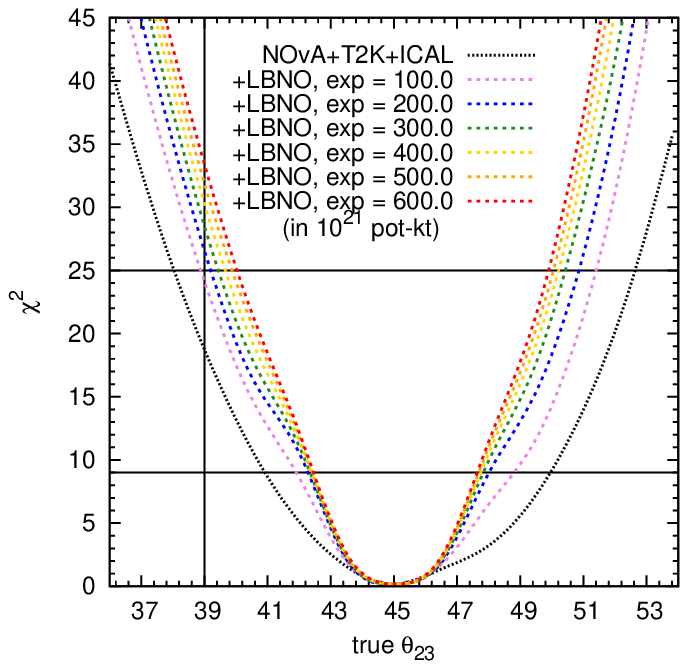,bbllx=85, bblly=50, bburx=300, bbury=250,clip=} &
\epsfig{file=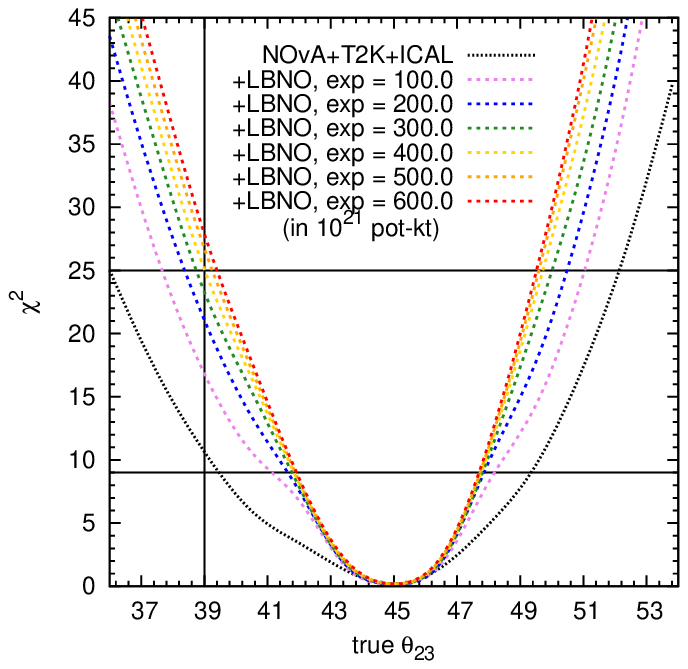,bbllx=85, bblly=50, bburx=300, bbury=250,clip=}
 \end{tabular}
\caption{\footnotesize Octant sensitivity $\chi^2$ vs true $\theta_{23}$ 
for the $130$ km baseline. The left(right) 
panel is for true NH(IH). In both panels, the lowermost densely-dotted (black) 
curve is for \nova+T2K+ICAL, while the curves above are for \nova+T2K+ICAL+LBNO, 
for various values of LBNO exposure. All the plotted sensitivities 
 are for the least favourable value of true $\dcp$.}
\label{fig:oct130}
\end{figure}

\begin{figure}[htb]
\begin{center}
\epsfig{file=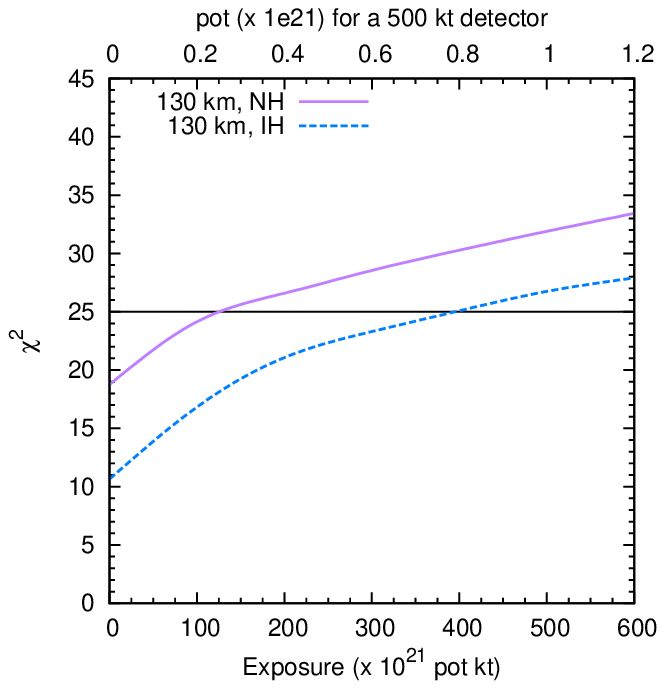, bbllx=85, bblly=50, bburx=300, bbury=255,clip=}
\end{center}
\caption{\footnotesize Octant sensitivity $\chi^2$ vs LBNO 
exposure, for the $130$ km baseline and both hierarchies, with 
$\theta_{23}=39^\circ$. The additional 
axis along the upper edge of the graph shows the required total pot assuming a 
detector mass of $500$ kt.}
\label{fig:oct130expo}
\end{figure}

\begin{figure}[htb]
\begin{tabular}{rl}
\epsfig{file=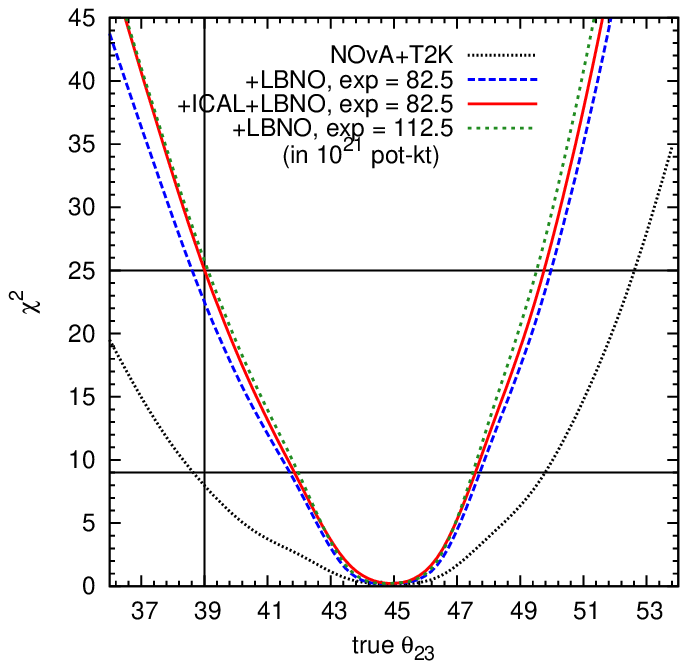, bbllx=85, bblly=50, bburx=300, bbury=255,clip=} &
\epsfig{file=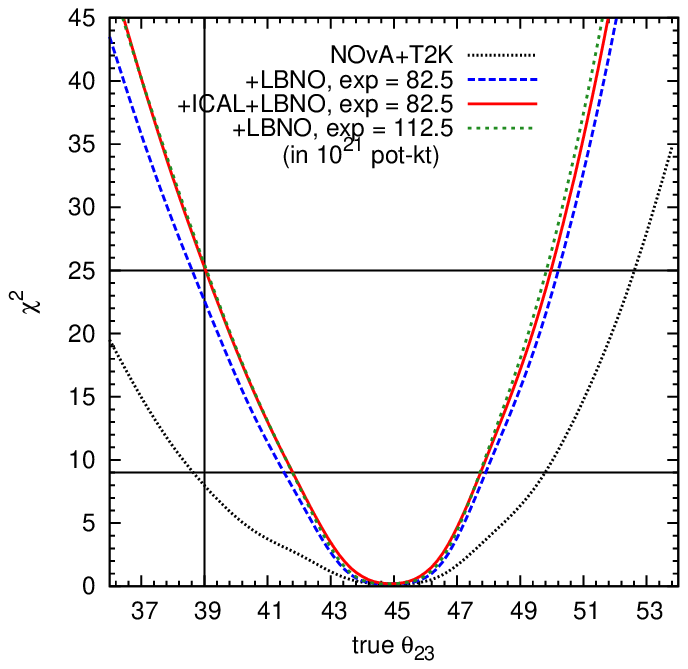, bbllx=85, bblly=50, bburx=300, bbury=255,clip=}
\end{tabular}
\caption{\footnotesize Octant sensitivity $\chi^2$ for different 
 combinations of experiments, demonstrating the synergy between them. The left(right) 
 panel is  for a LBNO baseline of $2290$($1540$) km, assuming IH to be true. 
 With only T2K+\nova+LBNO 
 (dashed, blue), the sensitivity is lower than for T2K+\nova+LBNO+ICAL (red, solid).
 Without ICAL data, the LBNO exposure would have to be increased substantially 
 (dotted, green) in order to get comparable sensitivity. All the plotted sensitivities 
 are for the least favourable value of true $\dcp$.}
\label{fig:octsyn}
\end{figure}

The octant sensitivity of long-baseline experiments has been studied in detail 
recently, both alone \cite{suprabhoctant,suprabhlbnelbno} and in conjunction 
with atmospheric neutrino experiments \cite{octant_atmos}. As in the case of 
hierarchy, adding information from various 
experiments enhances the sensitivity. However, it is the precise knowledge of 
the value of $\theta_{13}$ that plays a crucial role in determining the octant 
correctly. In Fig.~\ref{fig:octall}, the lowermost densely-dotted (black) 
curve denotes the ability of \nova+T2K+ICAL to determine the octant as 
a function of the true value of $\theta_{23}$ in nature. Again, the other curves 
denote the combined 
sensitivity of \nova+T2K+ICAL+LBNO for various values of LBNO exposure (in units of 
$10^{21}$ pot-kt). We generated the results for various 
true values of $\dcp$, and the 
results shown in the figure are for the most conservative case. We see that 
only with \nova+T2K+ICAL, the octant can be determined at $>3 \sigma$ C.L. 
when $\theta_{23}=39^\circ$. For values closer to $45^\circ$, the sensitivity 
gets steadily worse. The addition of LBNO data increases the sensitivity. For 
the range of exposures considered, it is possible to get a $\chi^2=25$ 
sensitivity to the octant as long as $\theta_{23}$ deviates from maximality 
by at least $\sim 6^\circ$. 

In 
Fig.~\ref{fig:octexpo}, we have shown how the octant sensitivity of these experiments 
increases as the exposure for LBNO is increased. For this, we have chosen the 
true value of $\theta_{23}$ to be $39^\circ$. 
Because of the better performance of \nova+T2K+ICAL when NH is true, the adequate 
exposure for LBNO is higher when IH is true. Given our current state of ignorance 
about the true hierarchy in nature, we list here the higher of the two numbers. 
It is sufficient to have an exposure of around 
$83\times10^{21}$ pot-kt to reach $\chi^2=25$ with both the baselines. 
The upper axis shows the total pot required, with a $10$ kt 
detector. For instance, we see that $8.3\times10^{21}$ pot is sufficient if we 
have a $10$ kt detector. This translates to a runtime of a little under 
$3$ years in 
each $\nu$ and $\overline{\nu}$ mode, given an intensity of $3\times10^{21}$ pot/yr. 

Fig.~\ref{fig:oct130} is the same as Fig.~\ref{fig:octall}, but for the $130$ km 
baseline. As expected, because of smaller matter effects, the exposure required 
to determine the octant is much higher than for the other two baselines. 
However, for a large mass detector like MEMPHYS that is being planned for the 
Fr\'{e}jus site, this exposure is not difficult to attain. The sensitivity as 
a function of LBNO exposure for this baseline is shown in Fig.~\ref{fig:oct130expo}. 
We need an exposure of around $400\times10^{21}$ pot-kt in this case. For this 
graph, the upper axis shows the required pot if we consider a $500$ kt detector, 
as proposed for MEMPHYS \cite{newmemphys}. We see that for such a large mass detector, 
only around $0.8\times10^{21}$ pot is adequate to exclude the octant for 
$\theta_{23} = 39^\circ$. Thus the beam intensity in pot is better than the 
other two set-ups. 

Fig.~\ref{fig:octsyn} shows the synergy 
between LBL experiments and ICAL. In the left(right) 
panel, we have chosen the LBNO baseline of $2290$($1540$) km to illustrate this 
point. IH is assumed to be the true hierarchy. The sensitivity of 
T2K+\nova\ alone (densely-dotted, black curve) is enhanced by adding data 
from ICAL and LBNO. The solid (red) curve in the left panel shows that an exposure 
of $82.5\times10^{21}$ pot-kt is enough to determine the octant with $\chi^2=25$ 
at $39^\circ$. But without ICAL data (dashed, blue curve), the sensitivity 
would be lower. The dotted (green) curve shows that only with $112.5\times10^{21}$ pot-kt
 (more than 35\% higher than the 
adequate amount), can we attain $\chi^2=25$ without ICAL. For $1540$ km 
(right panel) also, 
similar features are observed. This demonstrates the 
advantage of adding atmospheric neutrino data.

\section{Evidence for CP Violation}

\begin{figure}[htb]
\begin{tabular}{rl}
{\footnotesize 2290 km, NH} \hspace{2cm} & \hspace{3cm}  {\footnotesize 2290 km, IH} \\
\epsfig{file=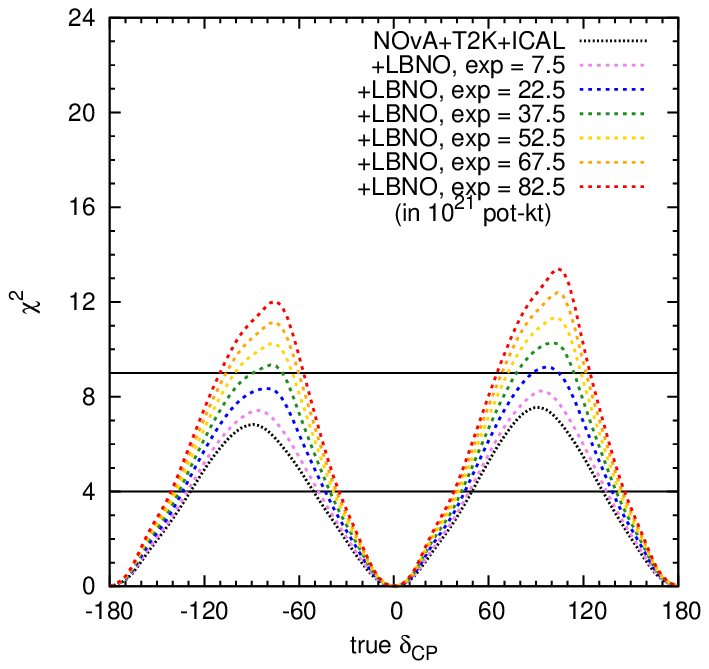,bbllx=85, bblly=50, bburx=300, bbury=250,clip=} &
\epsfig{file=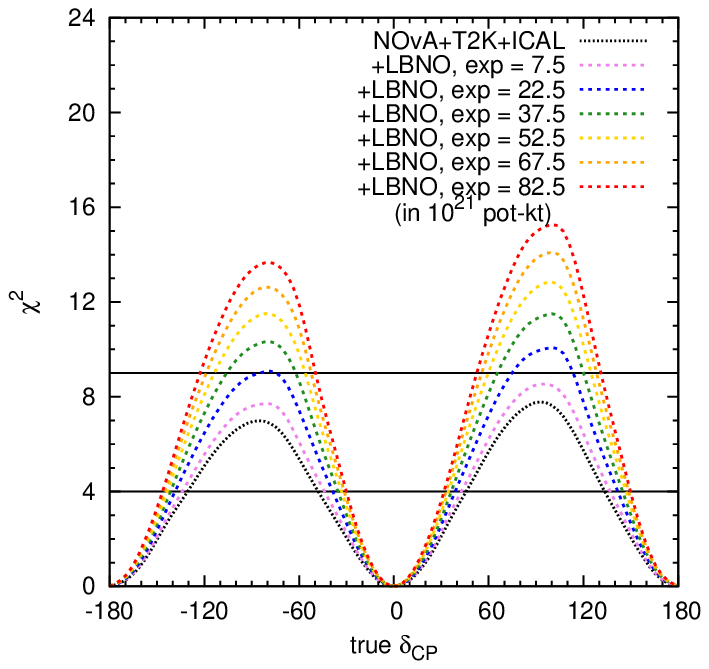,bbllx=85, bblly=50, bburx=300, bbury=250,clip=} \\
{\footnotesize 1540 km, NH} \hspace{2cm} & \hspace{3cm} {\footnotesize 1540 km, IH} \\
\epsfig{file=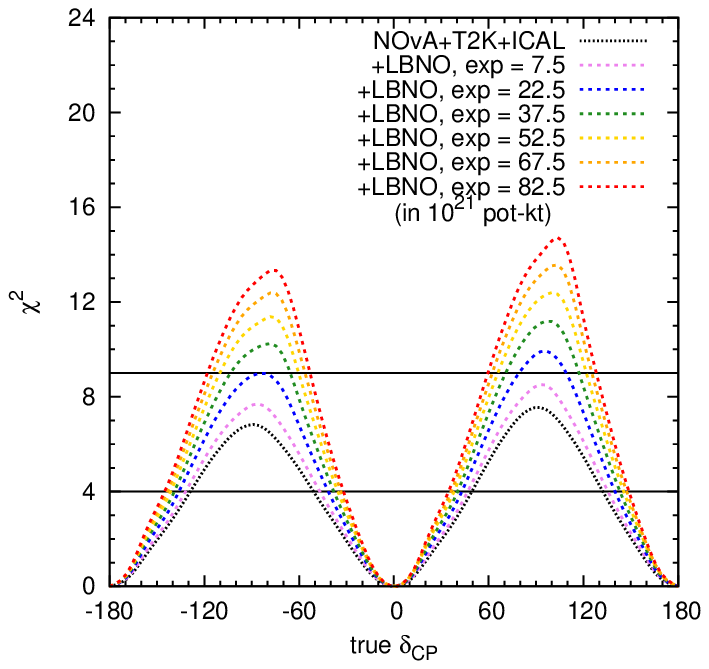,bbllx=85, bblly=50, bburx=300, bbury=250,clip=} &
\epsfig{file=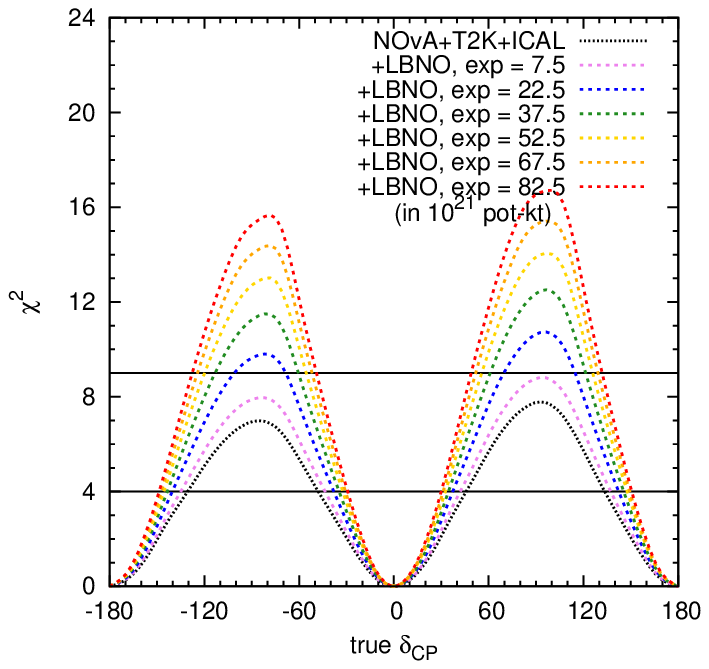,bbllx=85, bblly=50, bburx=300, bbury=250,clip=}
\end{tabular}
\caption{\footnotesize CP violation detection $\chi^2$ vs true $\dcp$. 
The top(bottom) panels are for the $2290$($1540$) km baseline. The left(right) 
panels are for true NH(IH). In all the panels, the lowermost densely-dotted (black) 
curve is for \nova+T2K+ICAL, while the curves above are for \nova+T2K+ICAL+LBNO, 
for various values of LBNO exposure. All the plotted sensitivities 
 are for the least favourable value of true $\theta_{23}$.}
\label{fig:cpdiscall}
\end{figure}

\begin{figure}[htb]
\begin{center}
\epsfig{file=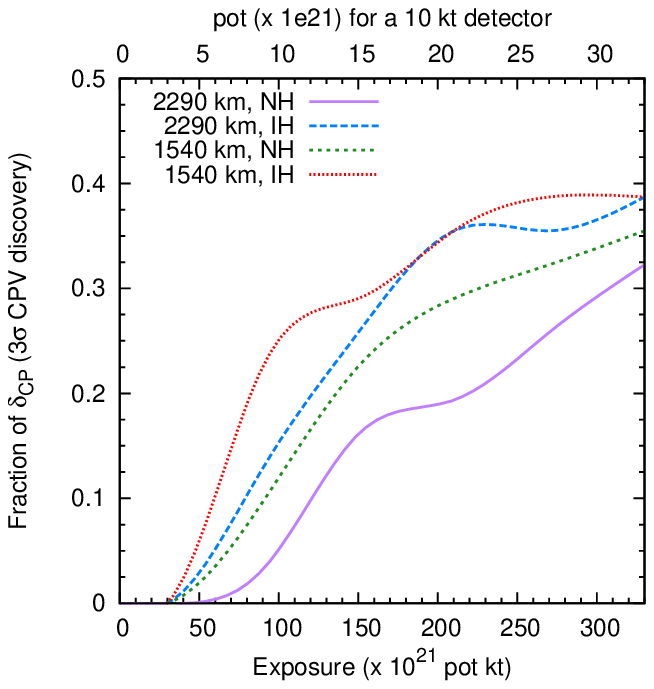, bbllx=85, bblly=50, bburx=300, bbury=255,clip=}
\end{center}
\caption{\footnotesize Fraction of the full $\dcp$ range for which it is 
possible to detect CP violation (exclude $\dcp=0,180^\circ$) at $3\sigma$ vs LBNO 
exposure, for the $2290$ km and $1540$ km baselines and both hierarchies. The 
additional 
axis along the upper edge of the graph shows the required total pot assuming a 
detector mass of $10$ kt.}
\label{fig:cpdiscexpo}
\end{figure}

\begin{figure}[htb]
 \begin{tabular}{rl}
  {\footnotesize 130 km, NH} \hspace{2cm} & \hspace{3cm} {\footnotesize 130 km, IH} \\
\epsfig{file=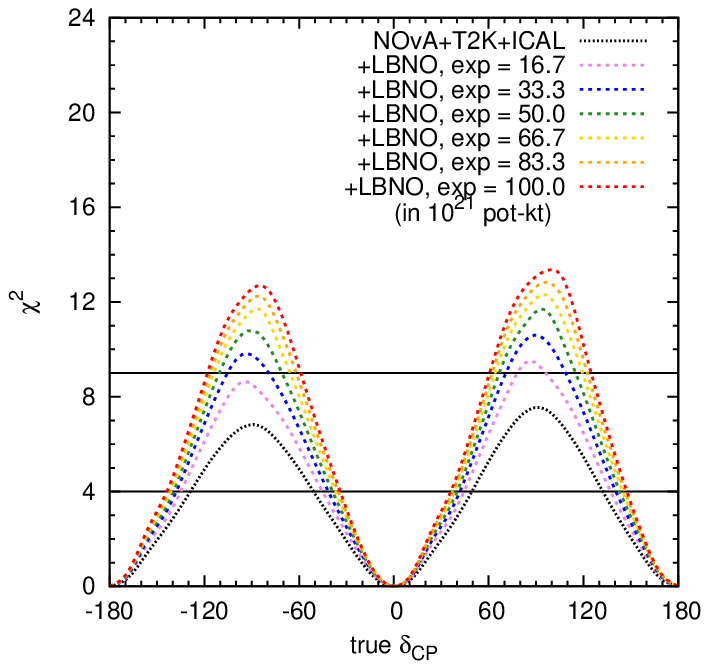,bbllx=85, bblly=50, bburx=300, bbury=250,clip=} &
\epsfig{file=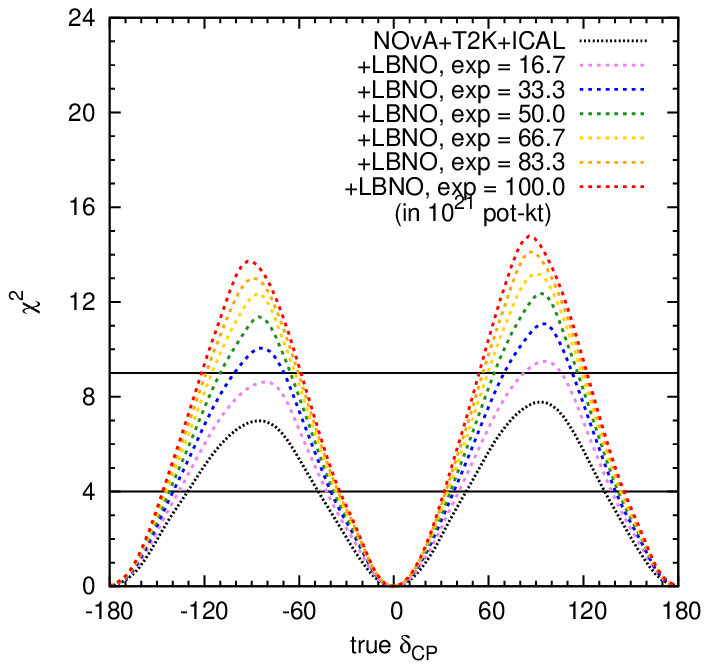,bbllx=85, bblly=50, bburx=300, bbury=250,clip=}
 \end{tabular}
\caption{\footnotesize CP violation detection $\chi^2$ vs true $\dcp$ 
for the $130$ km baseline. The left(right) 
panel is for true NH(IH). In both panels, the lowermost densely-dotted (black) 
curve is for \nova+T2K+ICAL, while the curves above are for \nova+T2K+ICAL+LBNO, 
for various values of LBNO exposure. All the plotted sensitivities 
 are for the least favourable value of true $\theta_{23}$.}
\label{fig:cpdisc130}
\end{figure}

\begin{figure}[htb]
\begin{center}
\epsfig{file=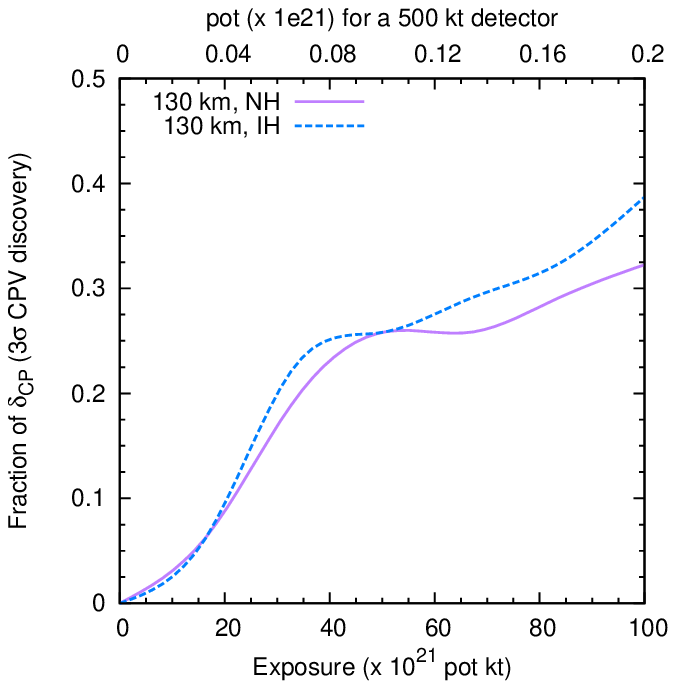, bbllx=85, bblly=50, bburx=300, bbury=255,clip=}
\end{center}
\caption{\footnotesize Fraction of the full $\dcp$ range for which it is 
possible to detect CP violation (exclude $\dcp=0,180^\circ$) at $3\sigma$ vs LBNO 
exposure, for the $130$ km baseline and both hierarchies. The additional 
axis along the upper edge of the graph shows the required total pot assuming a 
detector mass of $500$ kt.}
\label{fig:cpdisc130expo}
\end{figure}

\begin{figure}[htb]
\begin{center}
\epsfig{file=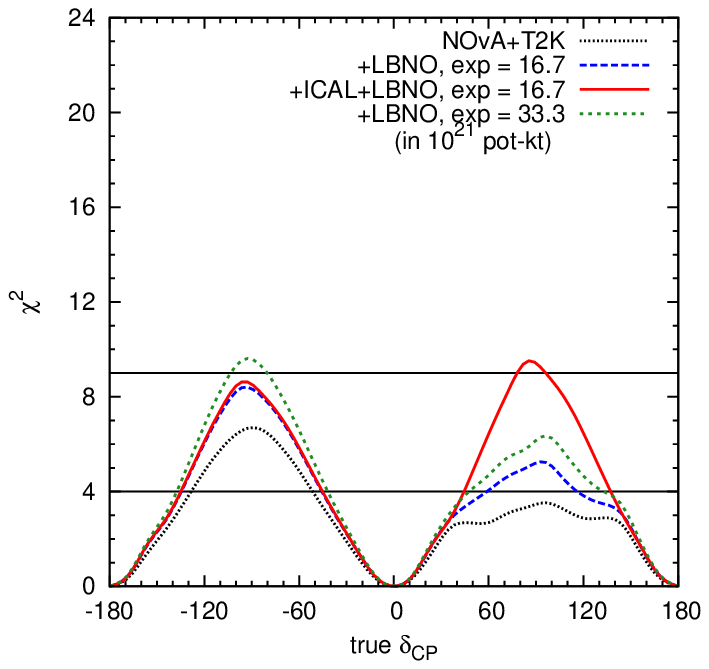, bbllx=85, bblly=50, bburx=300, bbury=255,clip=}
\end{center}
\caption{\footnotesize CP detection sensitivity $\chi^2$ for different 
 combinations of experiments, demonstrating the synergy between them. This plot is 
 for a LBNO baseline of $130$ km, assuming NH to be true. With only T2K+\nova+LBNO 
 (dashed, blue), the sensitivity is lower than for T2K+\nova+LBNO+ICAL (red, solid).
 Without ICAL data, the LBNO exposure would have to be increased substantially 
 (dotted, green) in order to get comparable sensitivity. All the plotted sensitivities 
 are for the least favourable value of true $\theta_{23}$.}
\label{fig:cpdiscsyn}
\end{figure}

Measurement of $\dcp$ is one of the most challenging problems in neutrino 
physics today. For the moderately large value of $\theta_{13}$ measured by 
the reactor neutrino experiments, it is possible for \nova\ and T2K to 
provide some hint on this parameter. In this paper, we discuss the detection
of CP violation, i.e. the ability of an experiment to exclude the cases 
$\dcp=0$ or $180^\circ$ 
\footnote{We emphasize that by `CP violation detection', we mean evidence 
that CP is violated in the neutrino sector. This is usually referred to in the 
literature as `CP violation discovery'. In this paper, we have avoided using 
this standard terminology, since the word `discovery' is usually taken 
to mean $5\sigma$ significance.}. 
We show our results as a function of $\dcp$ in 
Fig.~\ref{fig:cpdiscall}. Like in the case of hierarchy exclusion, we have 
minimized over three different true 
values of $\theta_{23}$, thus choosing the most 
conservative case possible. \nova\ and T2K suffer from the hierarchy-$\dcp$ 
degeneracy, because of which their CP detection potential is compromised 
for unfavourable values of $\dcp$. This degeneracy can be lifted by including 
information from ICAL, which excludes the wrong hierarchy 
solution \cite{cpv_ino}. Thus, in spite of not having 
any intrinsic $\dcp$ sensitivity, addition of atmospheric neutrino data 
improves the CP sensitivity of LBL experiments, provided 
the experiment itself does not have sufficient hierarchy sensitivity. 

We see in Fig.~\ref{fig:cpdiscall} that with \nova+T2K+ICAL, only around 
$\chi^2=4$ can be attained, for a small range of $\dcp$ values 
around $\pm 90^\circ$. Adding LBNO data with increasing exposure can enhance 
this, and even help to achieve $\chi^2=9$ CP detection for some range of 
$\dcp$. In Fig.~\ref{fig:cpdiscexpo}, we have plotted the fraction of $\dcp$ 
for which CP violation can be detected with $\chi^2=9$, as a 
function of the LBNO exposure. As an example, if we aim to detect CP 
violation for at least 20\% of $\dcp$ values, then we require around 
$240\times10^{21}$($170\times10^{21}$) pot-kt exposure from LBNO with a 
baseline of $2290$($1540$) km. 
It can also be seen from the figure that with $350\times10^{21}$ pot-kt exposure,
the maximum CP fraction for which a $3\sigma$ sensitivity is achievable 
ranges from 30\% to 40\%. 
The upper axis shows that these values correspond 
to $24\times10^{21}$($17\times10^{21}$) pot, if we consider a $10$ kt detector. 

Figs.~\ref{fig:cpdisc130} and \ref{fig:cpdisc130expo} show the results for 
the $130$ km option. Once again, we see that an exposure much higher than 
the longer baselines is required. In this case, CP detection for 20\% $\dcp$ 
values requires an exposure of around $35\times10^{21}$ pot-kt. This 
is not difficult to achieve with a large MEMPHYS-like detector. In fact, 
the total pot required by a $500$ kt detector at $130$ km is only around 
$0.07\times10^{21}$ pot. 
Moreover, an underground megaton scale detector like MEMPHYS can 
also be used to collect atmospheric neutrino data, which will further 
enhance the sensitivity  \cite{campagne}. 

In Fig.~\ref{fig:cpdiscsyn}, we have demonstrated the synergy between 
atmospheric and long-baseline experiments for the baseline of $130$ km 
and with NH. We see that with only T2K+\nova\ 
(densely-dotted, black curve), we suffer from the hierarchy-$\dcp$ 
degeneracy in the unfavourable region of $\dcp$. This degeneracy is lifted by 
adding information from other experiments. On adding data from ICAL and 
$16.7\times10^{21}$ pot-kt of LBNO (solid, red curve), we just reach 
$\chi^2=9$ sensitivity. 
With the same LBNO exposure, absence of ICAL data reduces the detection 
reach, as seen from the dashed (blue) curve. Reaching $\chi^2=9$ without 
ICAL will require the LBNO exposure to be doubled, as the 
dotted (green) curve shows. Thus, in spite of not having much intrinsic 
CP sensitivity, ICAL data contributes substantially towards CP sensitivity. 
For the two longer baselines, LBNO even with very low exposure in 
conjunction with T2K and \nova\ can break the hierarchy-$\dcp$ degeneracy 
by excluding the wrong hierarchy solution.
Therefore, the contribution of ICAL towards detecting CP violation 
becomes redundant in this case.

\section{Conclusion}

The reactor neutrino experiments have measured the value of $\theta_{13}$
to be moderately large. This is expected to facilitate the determination of the
three unknowns in neutrino oscillation studies --
the mass hierarchy, octant of $\theta_{23}$ and $\dcp$.
However the current LBL experiments T2K and \nova\ have limited
sensitivity to these parameters even for such large values of
$\theta_{13}$.  Combining the data from these experiments with
atmospheric neutrino data can result in
an enhanced sensitivity due to the synergistic aspects amongst them.
However a conclusive 5$\sigma$ evidence
would still be difficult to achieve and many future proposals are being
discussed for realizing this.

One of the most propitious among these
is the LBNO project in Europe. The exact design and baseline for
this is still under consideration.
In this paper we have explored the minimum exposure needed
for such a set-up and quantified  the
`adequate' configuration that can
exclude the wrong hierarchy
($\chi^2=25$), exclude the wrong octant ($\chi^2=25$) and detect CP violation
($\chi^2=9$). We have determined the adequate exposure required
for LBNO in units of pot-kt and for the least favourable 
true hierarchy, $\theta_{23}$ and $\delta_{CP}$. 
In determining the requisite exposure
we fully exploit the possible synergies between the
existing LBL experiment T2K, the upcoming LBL experiment
\nova\ and the  atmospheric neutrino
experiment ICAL@INO which is likely to commence data taking
in five years time.
For the prospective LBNO configuration we consider
three options:
CERN-Pyh\"{a}salmi ($2290$ km) baseline
with a LArTPC, CERN-Slanic
($1500$ km) with a LArTPC
and CERN-Fr\'{e}jus ($130$ km) with a Water \v{C}erenkov detector.
The `adequate' exposure needed is summarized
in Table~\ref{tab:res} where we give the results for T2K+\nova+LBNO
with and without ICAL. 
Inclusion of the atmospheric data from ICAL can play a significant role in
reducing the exposure required for hierarchy and octant determination
for the 2290 and 1540 km set-ups and for octant and CP detection for the
130 km set up.

\begin{center}
\begin{table}[tbh]
 \begin{tabular}{||l||c|c|c||}
  \hline
  \hline
  & \multicolumn{3}{c}{adequate exposure (pot-kt) for} \\
  & $2290$ km & $1540$ km & $130$ km \\
  \hline
  \hline
  Hierarchy exclusion ($\chi^2=25$) & $ 7(11) \times 10^{21}$ & $ 21(37) \times 10^{21}$ & $-$ \\
  \hline
  Octant exclusion at $39^\circ$ ($\chi^2=25$) & $ 83(113) \times 10^{21}$ & $ 83(113) \times 10^{21}$ & $ 400(600) \times 10^{21}$ \\
  \hline
  \parbox[c]{0.35\columnwidth}{CP violation detection ($\chi^2=9$) \\for 20\% fraction of $\dcp$} & $ 240(240) \times 10^{21}$ & $ 170(170) \times 10^{21}$ & $ 35(100) \times 10^{21}$ \\
  \hline
  \hline  
 \end{tabular}
 \caption{\footnotesize Summary of results: `adequate' exposure in pot-kt 
 for three LBNO configurations to achieve the physics goals. The numbers given in 
 parentheses indicate the required exposure if atmospheric neutrino data from ICAL 
 is not included.}
 \label{tab:res}
\end{table}
\end{center}

Of the two longer baselines, we find that $2290$ km is best 
suited to determine the mass
hierarchy, while $1540$ km is better for detecting CP violation.
However, $130$ km is the best candidate for CP violation physics.
The `adequate' exposures listed in this work can be attained by various
combinations of beam power, runtime and detector mass. These minimal values can
be used to set up the first phase of LBNO, if an incremental/staged approach is
being followed.
Finally, we would like to emphasize that
the synergies between the existing and upcoming
LBL and atmospheric experiments can play an
important role and should be taken into consideration in
planning economised future facilities.

\subsection*{Acknowledgements}

S.R. would like to thank Sanjib Agarwalla for information on experimental details; 
Robert Wilson for useful discussions; and the neutrino 
physics group at Laboratoire A.P.C., Universit\'{e} Paris Diderot for their 
hospitality during the writing of this manuscript. 

\bibliography{neutosc}

\end{document}